\documentclass[12pt]{iopart}

\usepackage{iopams}
\usepackage{graphics}

\newcommand{\react}[1]{\stackrel{#1}{\longrightarrow}}
\newcommand{\ave}[1]{\langle #1\rangle}
\newcommand{\eq}[1]{Eq.\,(\ref{#1})}
\newcommand{\eqs}[2]{Eqs.\,(\ref{#1},\ref{#2})}
\newcommand{\der}[1]{\partial_{#1}}
\newcommand{\Der}[1]{\frac{\partial}{\partial{#1}}}

\begin{document}

\title[Connecting protein and mRNA burst distributions]{Connecting protein and mRNA burst distributions for stochastic models of gene expression}

\author{Vlad Elgart\footnote{Present Address: Department of Microbiology and Immunology, Baxter Lab for Stem Cell Biology, Stanford University, School of Medicine, Palo, Alto, California}, Tao Jia, Andrew T.\ Fenley and Rahul Kulkarni}

\address{Department of Physics, Virginia Tech, Blacksburg, VA 24061}
\ead{\mailto{elgart@vt.edu}, \mailto{kulkarni@vt.edu}}


\begin{abstract}
The intrinsic stochasticity of gene expression can lead to large
variability in protein levels for genetically identical cells. Such
variability in protein levels can arise from infrequent
synthesis of mRNAs which in turn give rise to bursts of protein
expression. Protein expression occuring in bursts has indeed been
observed experimentally and recent studies have also found evidence
for transcriptional bursting, i.e.\ production of mRNAs in bursts.
Given that there are distinct experimental techniques for quantifying
the noise at different stages of gene expression, it is of interest to
derive analytical results connecting experimental observations
at different levels.  
In this work, we consider stochastic models of 
gene expression for which mRNA and protein production occurs in 
independent bursts. 
For such models,
we derive analytical expressions connecting
protein and mRNA burst distributions which show how the functional
form of the mRNA burst distribution can be inferred from the protein
burst distribution.  Additionally, if gene expression is repressed
such that observed protein bursts arise only from single mRNAs, we
show how observations of protein burst distributions (repressed and
unrepressed) can be used to completely determine the mRNA burst
distribution. Assuming independent contributions from individual
bursts, we derive analytical expressions connecting means and variances for burst and
steady-state protein distributions. Finally, we validate our general analytical
results by considering a specific reaction scheme involving regulation
of protein bursts by small RNAs. For a range of parameters, we derive
analytical expressions for regulated protein distributions that are
validated using stochastic simulations. The analytical results obtained in 
this work can thus serve as useful inputs for a broad range of
studies focusing on stochasticity in gene expression.

\end{abstract}

\submitto{\PB}


\maketitle

\section{Introduction}

The intrinsic stochasticity of biochemical reactions has important
consequences for the functioning of cellular processes
\cite{kaern05,paulsson05}.  In particular, reactions corresponding to
the process of gene expression often involve small numbers of
molecules, and can be subject to large fluctuations. The corresponding
stochasticity in gene expression has been identified as a key factor
underlying the observed phenotypic variability of genetically
identical cells in homogeneous environments \cite{raj08}. Quantifying
the effects of intrinsic noise using stochastic models of gene
expression is thus an important step towards understanding cellular
function and variability.

Several recent studies have focused on quantifying noise in gene
expression using both single-cell assays and single-molecule
techniques.  Experimental observations of noise in steady-state
protein distributions across a population of cells \cite{ozbudak02}
were shown to be consistent with predictions from simple models based
on translation from individual mRNAs \cite{ozbudak02,thattai01}. These
models predict that each mRNA produces a burst of protein that is
geometrically distributed \cite{berg78}.  Single-molecule studies have
indeed seen protein production occurring in bursts and determined that
the corresponding protein burst distribution is geometric
\cite{yu06,cai06,taniguchi}. At the mRNA level, single-molecule studies have
demonstrated that mRNA production can also occur in transcriptional
bursts \cite{raj08,yu06,golding05,raj06,chubb06,kaufmann07}. The
presence or absence of transcriptional bursting indicates different
sources of noise in gene expression and several
studies are currently engaged in probing gene expression at multiple
stages to elucidate the underlying sources of variability 
\cite{kaufmann07,azaele09,elgart_pre}.

Given different experimental techniques for probing stochasticity in
gene expression using measurements at different stages (specifically
steady-state and burst distributions for proteins and mRNAs)
\cite{raj09,larson09}, it is of interest to derive analytical results
connecting observables at different levels. These results can be used
to infer information at one level using experiments at a different
level. For example, in previous work \cite{friedman06} it was shown
that experimental determination of the protein burst distribution and
frequency can be used to determine the steady-state protein
distribution. In this context, we note that most previous models have
focused on reaction schemes which correspond to a geometric burst
distribution for proteins produced from a single mRNA
\cite{friedman06,ingram08}. However, more general reaction schemes for
protein production from mRNAs can lead to deviations from geometric
burst distributions for single mRNA bursts \cite{jia10}. 
It would thus be desirable
to derive analytical formulae connecting burst and steady-state
protein distributions for arbitrary protein burst
distributions. Finally, we note that such analytical results can be
used to check for consistency between the experimental results from
probing different levels of gene expression. In particular, any
observed inconsistencies could signal that some model assumptions are
invalid, potentially leading to new insights about the mechanisms of
gene expression.

In this work, we analyze a class of burst models for protein production from
mRNAs and derive analytical results connecting observable
distributions at different stages of gene expression. In particular, we show how the
functional form of the mRNA burst distribution can be determined using
the observed protein burst distribution.  If mRNA transcription can be
repressed such that observed protein bursts arise only from single
mRNAs, then the derived results show how observations of protein burst
distributions (repressed and unrepressed) can be used to completely
determine the mRNA burst distribution.  Assuming independent bursts
whose arrival can be modeled as a Poisson process, we derive
expressions connecting the mean and variance of protein burst distributions to the
corresponding quantities for the steady-state distribution of protein levels across a
population of cells. Finally, we consider a specific example for which
burst distributions can deviate from the geometric distribution:
post-transcriptional regulation of bursts by small RNAs. In the limit
of low burst frequency, we derive analytical expressions for the
protein burst distribution which are in excellent agreement with
results from stochastic simulations. The results derived in this work
can thus serve as useful building blocks for future studies focusing
on stochasticity in gene expression.

\section{Tools, Notations, and Definitions}

A starting point of our analysis is the Master equation \cite{kampen}
\begin{eqnarray}
        \der{t}P(\vec n;\;t) = \hat H(\vec n)\, P(\vec n;\;t),\label{master}
\end{eqnarray}
where $\vec n = \{n_X\}$ is a state vector describing the abundance of
each species $X$ in the system. Here $P(\vec n;\;t)$ is the
probability to find the system with the state vector $\vec n$ after
time $t$ has elapsed. 
Equation (\ref{master})
is supplemented by the initial conditions, namely the initial
distribution $P_0(\vec n)$ at time $t=0$.

\noindent 
The generating function of the probability
distribution \eq{master} is defined by
\begin{eqnarray}
        G(\vec x;\; t)\equiv \sum_{n_i=0}^\infty P(\vec n;\; t)
        \prod_{i=\{X\}} x_i^{n_i},
\end{eqnarray}
where $\vec x = \{x_i\}$ is a real vector dual to the state vector $\vec n$.
The generating function, in turn, satisfies the corresponding evolution
equation and initial condition
\begin{eqnarray}
        \der{t}G(\vec x;\; t) = \hat{\cal H}(\vec x)\, G(\vec x;\; t),\\
        G(\vec x;\; 0) = G_0(\vec x) \equiv
        \sum_{n_i=0}^\infty P_0(\vec n) \prod_i x_i^{n_i}.
\end{eqnarray}

\begin{figure}[h!]
\begin{center}
\resizebox{75mm}{!}{
\includegraphics{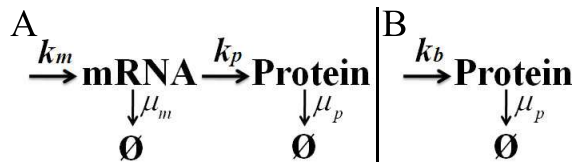}}
\caption{Reaction scheme for the minimal model of gene expression.
\emph{A.} mRNA transcripts are created with reaction rate $k_m$ and
degraded with reaction rate $\mu_m$. Protein is translated from mRNA
with a reaction rate $k_p$ and degraded with a reaction rate $\mu_p$.
\emph{B.} Evolution of protein distributions for the same scheme can
be analyzed in terms of arrivals of bursts of proteins (valid
when $\mu_p << \mu_m$) \cite{friedman06}.}
\label{figure_min_scheme}
\end{center}
\end{figure}

Let us first consider the simplest gene expression reaction scheme.  The
minimal model \cite{berg78} of gene expression is given by the diagram
on Fig.\ref{figure_min_scheme}.  The corresponding reaction
scheme is

\begin{equation}
	D\react{k_m}{M}\react{k_p}{M+P};\quad M\react{\mu_m}\emptyset;
	\quad P\react{\mu_p}\emptyset; \label{scheme-min}
\end{equation}

\noindent
where $D$ is DNA, $M$ is mRNA, and $P$ is protein.
Both mRNAs and proteins are synthesized at the constant rates $k_m$
and $k_p$ respectively, and their degradation (decay) rates are
$\mu_m$ and $\mu_p$.
Correspondingly, the evolution operator in the equation for the generating 
function is given by
\begin{eqnarray}
       \hat{\cal H} = k_m(x_m - 1) + \mu_m(1 - x_m)\Der{x_m} + \nonumber\\
       k_p(x_p - 1)x_m\Der{x_m} + \mu_p(1 - x_p)\Der{x_p}.
\end{eqnarray}
Note that generating function representation is particularly useful
since it converts the infinite set of equations for various integer values of 
$\vec n$ in \eq{master} into a single partial differential
equation. Moreover, a calculation of any observable quantities, such
as moments of distribution, is equivalent to evaluation of derivatives
of the generating function at point $\vec x=\{1,1,\cdots,1\}$.

There are several parameters that describe the dynamics of the gene
expression models analyzed in this work. The following rules serve as general 
guides for the notation used:

\begin{itemize}
\item Indices $X=m,p,s$ stand for mRNA, protein, and sRNA species correspondingly.
\item Lower case letters are used to describe burst variable, e.g., $p_m$ denotes the probability distribution of mRNA burst size and $g_m$ is its 
generating function.
\item Capital letters are used to describe steady-state variable, e.g., $P_p$ denotes the protein steady-state distribution and $G_p$ is its generating function.
\end{itemize}

Finally, we define some distributions that arise when considering bursts of
gene expression. The geometric distribution is given by
\begin{equation}
        \tilde{\rho}(n) = (1-u)^{n}u ,\quad n\ge 0\label{geom-dist}
\end{equation}
with the corresponding generating function
\begin{eqnarray}
        \tilde{G}(x) = \frac{u}{1 - (1-u)\,x}.
\end{eqnarray}
and mean given by $(1-u)/u$. It is also convenient to define
the {\em conditional} geometric distribution
\begin{eqnarray}
        \rho(0) = 0,\nonumber\\
        \rho(n) = (1-u)^{n-1}u,\quad n\ge 1\label{cond-geom-dist}
\end{eqnarray}
with the corresponding generating function
\begin{eqnarray}
        G(x) = \frac{u\, x}{1 - (1-u)\,x}.
\end{eqnarray}

The conditional geometric distribution is encountered \cite{ingram08} 
when considering the distribution
of mRNAs that give rise to a protein burst. Since the observation of a
burst of proteins necessarily implies the presence of at least 1 mRNA,
the distribution is conditioned accordingly. We note that the mean of the
conditional
distribution is given by $1/u$. In the limit
$u\rightarrow 1$, the distribution \eq{cond-geom-dist} describes a
mRNA burst with exactly 1 mRNA produced per burst, which is the case
when mRNA arrival corresponds to a Poisson process.

\section{Bursts and modeling framework}\label{sect-BA}

\begin{figure}[h!]
\begin{center}
\resizebox{119.63mm}{!}{
\includegraphics{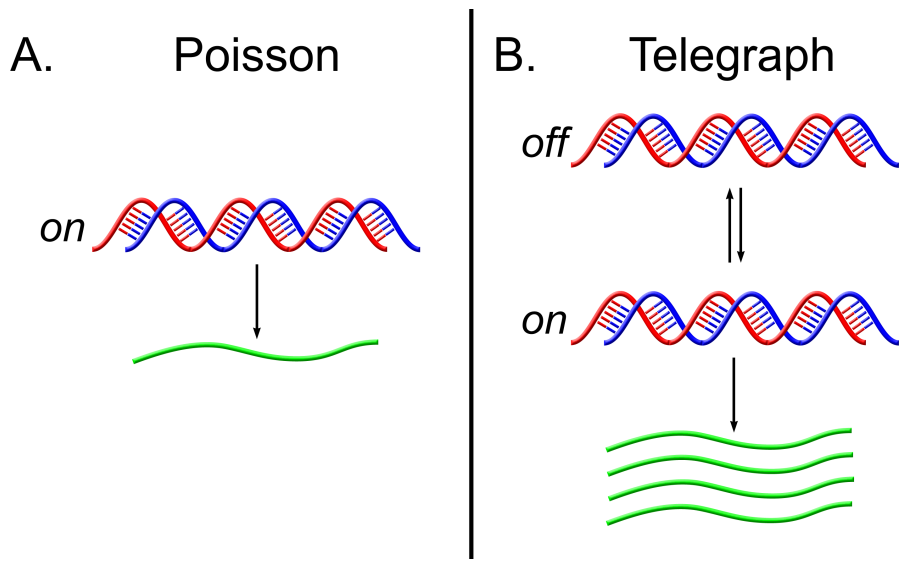}}
\caption{(color online) Schematic representation of the Poisson and
Telegraph mRNA transcription processes. \emph{A.}\ The Poisson process
of transcription.  The DNA is always in the ``on" state resulting in a
constant production rate of mRNA transcript (green lines). \emph{B.}
The Telegraph process of transcription. The DNA exchanges between two
states, ``off" and ``on".  The ``off" state corresponds to inactive DNA
in which no transcripts are produced, while the ``on" state corresponds
to DNA capable of producing a burst of mRNA transcripts before it
reverts back to the ``off" state.}
\label{figure_telegraph_poisson}
\end{center}
\end{figure}

Recent experiments have determined the variation of noise in protein
expression as a function of mean protein abundance for several genes
\cite{bareven06, newman06}. The observed scaling relationship is
consistent with different underlying models
(see Figure \ref{figure_telegraph_poisson}).
In one case, the transcription rate $k_m$ is constant corresponding to 
a Poisson process driving mRNA synthesis.  Another possible scenario 
corresponds to a Telegraph process 
\cite{paulsson05,golding05,karmakar04,iyer09,bruggeman09} 
driving the
creation of mRNAs. In this case, the promoter driving gene expression
switches between active and inactive states. When the promoter is in
the active state, multiple number of mRNAs can be
transcribed. While both models are consistent with the experimental data, the
observed scaling indicates that protein production occurs in
infrequent bursts for many genes.

Based on observations relating to bursts, an analytical
approach \cite{friedman06} was introduced to derive expressions for
steady-state protein distributions from protein burst
distributions. Specifically, it is assumed that (i) protein degradation rate 
is much smaller than mRNA degradation rate ($\mu_p \ll \mu_m$), (ii) protein 
levels
vary due to {\it independent} bursts of protein expression in
combination with changes due to protein degradation and (iii) the arrival
of bursts can be modeled as a Poisson process.  The above approach
then reduces the problem of characterizing protein steady-state
distributions into two parts: (i) first obtain the protein burst
distribution for a single burst and (ii) using this burst distribution
as input, derive and analyze the corresponding Master equation (see Fig. 1B)
for proteins alone \cite{friedman06,Shahrezaei2008}.  A mathematical
justification of this procedure of deriving a Master equation for
proteins alone, given the assumptions stated above, has been provided
recently \cite{Shahrezaei2008}.

\begin{figure}[h!]
\begin{center}
\resizebox{124mm}{!}{
\includegraphics{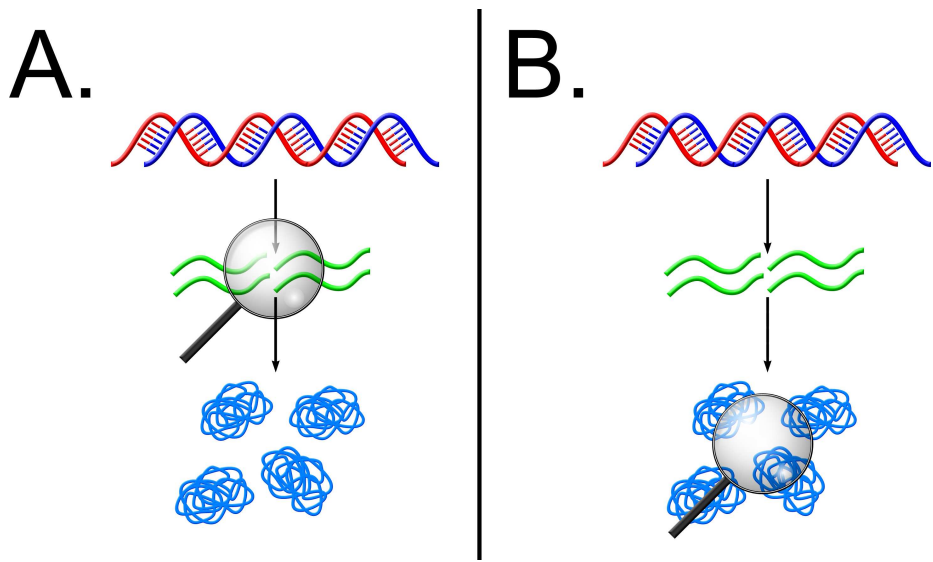}}
\caption{Different approaches for probing bursts of gene expression.
\emph{A.} Measuring the mRNA burst distributions directly.
\emph{B.} Measuring the protein distributions resulting from the mRNA burst
distributions. The derived results provide means of connecting these two
measurements at different stages of gene expression for the case of arbitrary protein burst
distributions arising from a single sRNA.}
\label{figure_mRNA_vs_Protein}
\end{center}
\end{figure}

In the following sections, we will consider stochastic models of gene
expression consistent with the assumptions stated above.  
Specifically, we consider models for which mRNA and protein production occurs 
in independent bursts such that the arrival of bursts corresponds to a Poisson
process. As noted above, even for a Poisson process, 
there are parameter constraints that must be satisfied 
($\mu_p \ll \mu_m$) for the burst approximation to be valid.
While previous work has largely focused on reaction schemes that give rise
to a geometric burst of proteins from a single mRNA, we will 
consider the case for which the protein burst distribution can be an 
arbitrary function.
For such models, we wish to derive analytical expressions
which connect observations at different stages of gene expression 
(see Figure  \ref{figure_mRNA_vs_Protein}). The following section first 
considers how observations of protein burst distributions can inform us about 
the underlying mRNA burst distributions.

\section{From protein to mRNA burst distribution}

We first consider the minimal scheme (\eq{scheme-min}) of protein
production from mRNAs. For this scheme, the following equation
relating the mRNA and protein burst distributions can be derived
(see Appendix):
\begin{eqnarray}
	g_m(x) = g_p\left[1 - \frac{(1-x)}{(k_p/\mu_m)x}\right],\label{gen-mp}
\end{eqnarray}

The functions $g_m$ and $g_p$ in the equation \eq{gen-mp} are
correspondingly the mRNA and protein burst generating functions. The
dynamical version (for time dependent distributions) of 
\eq{gen-mp} can also be found in the Appendix. Note that, consistent
with the assumption $\mu_m\gg\mu_p$, we ignore protein degradation
during a single burst, i.e.\ the above equation considers only the
proteins synthesized during the burst.

The result \eq{gen-mp} is useful because it allows us to infer the
functional form of the mRNA burst distribution from observations of
protein burst distributions. Consider the case that the observed protein 
burst distribution (e.g.\ as reported in \cite{cai06}) is a geometric 
distribution (\eq{geom-dist}) with parameter $u =u_p$. Then, using the 
expression 
\eq{gen-mp}, we obtain that the mRNA burst distribution has to be a 
conditional geometric distribution, \eq{cond-geom-dist}, with parameter
\begin{equation}
	u_m = \frac{k_p}{\mu_m}\frac{u_p}{1-u_p}.
\end{equation}
In other words, the mRNA burst distribution is given by
\begin{eqnarray}
	P_m(n) = \rho(u_m),\quad n\ge 1\label{mRNA-dist}
\end{eqnarray}

While the functional form of the mRNA burst distribution is thus
determined, we note that the precise distribution is not known since
the parameter $(\frac{k_p}{\mu_m})$ is not known.  The upper bound for
$\frac{k_p}{\mu_m}$ can be derived from the condition $u_m =1$
\begin{equation}
 \left(\frac{k_p}{\mu_m} \right)_{\mathrm{max}} =\frac{1-u_p}{u_p},
\end{equation}
which corresponds to the Poisson scenario, i.e. the observed burst
distribution is produced from a single mRNA. On the other hand, we can
have $(\frac{k_p}{\mu_m}) < (\frac{k_p}{\mu_m})_{\mathrm{max}}$, which 
implies $u_m < 1$ 
and thereby that the mean number of mRNAs in the
burst ($\frac{1}{u_m}$) is greater than 1.  This set of parameters would 
be consistent with a Telegraph process driving mRNA creation since it
produces a geometric mRNA burst distribution (with $u_m < 1$) and also
gives rise to a geometric protein burst distribution \cite{ingram08}.

It has been noted in previous work \cite{golding05,bareven06,ingram08}
that the Poisson and Telegraph processes cannot be distinguished 
by experimental observations on proteins alone, since both
both Poisson and Telegraph processes give
rise to a geometric burst distribution for proteins.  However,
previous work did not preclude other possible mRNA burst distributions
that can result in a geometric protein burst distribution. The preceding
analysis demonstrates that, if the observed protein burst
distribution is geometric, then  the mRNA burst distribution has to be a
conditional geometric distribution. 
Thus \eq{mRNA-dist} is a mathematically
necessary and sufficient condition on the mRNA burst distribution to
obtain a geometric burst distribution for proteins.  An important corollary  
is that
kinetic schemes which lead to non-geometric mRNA burst distributions
can be ruled out if the observed protein burst distribution is a
geometric distribution.

Let us now consider general reaction schemes which can give rise to
non-geometric protein burst distributions. This can occur due to
interaction with a post-transcriptional regulator \cite{jia10} or even 
otherwise,
e.g.  if we have switching between competing mRNA secondary structure
conformations which correspondingly have different protein production
and/or mRNA degradation rates.  Another example is the case for which 
mRNA degradation is not a Poisson process but occurs in stages (termed 
mRNA senescence \cite{pedraza08}); in general the corresponding protein
burst distribution will not be a geometric distribution. In such cases, the
preceding analysis can be generalized as follows.  Let us denote by
$\phi(x)$ the generating function of protein bursts obtained from a
{\em single} mRNA. The number of proteins produced in a single burst
can be expressed as the sum of a random number ($N$) of random
variables, each of which is drawn from the probability distribution
corresponding to $\phi(x)$. The random variable $N$ corresponds to the
number of mRNAs in the burst with generating function
$g_{m}(x)$. Correspondingly, the generating function of the protein burst
distribution is given by
\begin{eqnarray}
        g_p(x) = g_m\left[\phi(x)\right],\label{g_p}
\end{eqnarray}
Inversion of  \eq{g_p} yields the mRNA burst distribution
\begin{eqnarray}
        g_m(z) = g_p\left[\phi^{-1}(z)\right],\label{g_m}
\end{eqnarray}
Note that \eq{gen-mp} is a special case of \eq{g_m}. For the minimal scheme 
of gene expression (Fig. 1), the burst distribution from a single mRNA 
is a geometric distribution with mean $\frac{k_p}{\mu_m}$ \cite{berg78}. 
Correspondingly, the generating function is given by $\phi(x) = 
\frac{u}{1- (1-u)x}$, with $u= \frac{\mu_m}{k_p + \mu_m}$. Inversion of 
$\phi(x)$ in combination with \eq{g_m} gives \eq{gen-mp}.

The significance of the above equations is that once $\phi(x)$ is determined,
the mRNA burst distribution can be inferred from the observed
protein burst distribution (and vice-versa). Recent experiments
\cite{cai06} have shown that repressors can be used to regulate gene
expression such that each observed burst corresponds to proteins
produced from a single mRNA. Such experiments can be used to determine
the single mRNA burst distribution and hence $\phi(x)$. Thus, if the
protein burst distributions can be observed for both scenarios, with and
without the repressor, then \eq{g_m} can be used to completely
determine the mRNA burst
distribution.

\section{Connecting Burst and Steady-State Distributions}\label{sect-GSF}

While the direct observation of protein expression bursts has been
demonstrated experimentally \cite{yu06,cai06,taniguchi}; in general, carrying
out such experiments is challenging. Since steady-state protein
distributions are less challenging to determine experimentally, it is
of interest to derive results connecting burst and steady-state
distributions, in particular connecting the means and variances.  
We note that recent work \cite{pedraza08} has derived results connecting 
burst and steady-state variances for general models of gene expression, in 
particular for models such that the waiting-time distribution between bursts 
can arbitrary, as opposed to the simple exponential distribution which 
corresponds to a Poisson process for burst arrival. In the following, we
first focus on the case of Poisson arrivals for bursts.

As discussed in Section \ref{sect-BA}, we assume that each burst can be
considered as an independent realization of the same stochastic
process and that burst arrival can be modeled as a Poisson process.
Let us denote by $P_b(n)$ the probability that $n$ proteins are produced during
a single burst. Correspondingly, the Master equation for the protein
distribution at time $t$ ($P(n,t)$) is \cite{IyerBiswas}:
\begin{eqnarray}
	\partial_t P(n,t) = \mu_p \left[P(n+1,t)-P(n,t)\right]\nonumber\\
	+ \,k_b \sum_{n'=0}^\infty\left[P_b(n')P(n-n',t)-P_b(n')P(n,t)\right]\label{burst-master}
\end{eqnarray}
The parameter $k_b$ is the constant rate of burst arrival, i.e. it is
the inverse of the mean time between two sequential bursts. If each burst
corresponds to proteins produced from a single mRNA, then $k_b$ is identical
to the mRNA creation rate $k_m$.

Let us define the generating functions:
\begin{eqnarray}
	G_{b}(x) = \sum_{n=0}^\infty x^n P_{b}(n),\\
	G(x,t) = \sum_{n=0}^\infty x^n P(n,t)
\end{eqnarray}
Correspondingly, the evolution equation for the generating function is \cite{IyerBiswas}
\begin{eqnarray}
	\partial_t G(x,t) = \mu_p(1-x)\partial_x G(x,t)
	+\,k_b \left[G_b(x)-1\right]G(x,t),\label{burst-gen}
\end{eqnarray}

The time dependent solution of \eq{burst-gen} can be obtained
by the method of characteristics. The steady-state limit
($G_{s}(x)$) is given by \cite{IyerBiswas}, 
\begin{equation}\label{steady}
	G_s(x) = \exp\left\{\frac{k_b}{\mu_p}\int_1^x \left(
\frac{G_b(y)-1}{y-1} \right) dy\right\}.
\end{equation}

From \eq{steady} we can also derive useful expressions for the mean and the
Fano factor (or noise strength) of the steady-state distribution in
terms of the corresponding quantities for burst distribution. We
obtain:
\begin{eqnarray}
	\bar{n}_s = \left(\frac{k_b}{\mu_p} \right) \bar{n}_b,\label{mean}\\
	\frac{\sigma^2_s}{\bar{n}_s} = 1 + \bar{n}_b + \frac{1}{2} \left( \frac{\sigma^2_b}{\bar{n}_{b}} - (1 + \bar{n}_b) \right) .\label{var}
\end{eqnarray}
If the burst distribution is geometric, we have
$\frac{\sigma^2_b}{\bar n_{b}} = 1 + \bar{n}_b $ and the above result
reduces to previously obtained results
\cite{paulsson05,Shahrezaei2008} in the limit $\mu_m \gg \mu_p$. For
general reaction schemes, the burst distribution differs from the
geometric distribution and \eq{var} is the generalization that
connects burst and steady-state distributions.  It is interesting to
note that a similar result was obtained in previous work
\cite{pedraza08} with different model assumptions: specifically, each
mRNA was assumed to produce a geometric burst of proteins, however the
number of mRNAs in the burst was assumed to be drawn from an arbitrary
burst distribution.

The preceding discussion focused on the case that the burst arrival is
a Poisson process, thus the waiting-time distribution between bursts
is given by an exponential distribution. For a Poisson process driving
mRNA production this is certainly the case. However, mRNA production
has also been proposed to arise from a Telegraph process
\cite{golding05,kaufmann07} which, in general, does not have the
feature that the waiting-time between bursts is an exponential
distribution \cite{bruggeman09}.  We consider the case that the DNA
fluctuates between two different conformations which correspond to
different production rates for the mRNAs. In particular, we consider a
two-stage model \cite{golding05} corresponding to two different active
confirmations of DNA (``on" and ``off" or 1 and 2 say) , with mRNA 
transcription rates $k_1$
and $k_2 (< k_1)$ respectively, Note that previous work has focused on
the case $ k_2 = 0$, i.e.\ no transcription in the ``off" state. The
present results generalize this model to allow for a basal level of
transcription in the ``off" state as well. Let us define a parameter
$f = \frac{k_2}{k_1}$ which is the ratio of the two rates and takes
values between $0$ and $1$.
We denote by
$\lambda_{12}$ the rate of switching from conformation $1$ to conformation
$2$, and $\lambda_{21}$ is the rate for the reversed process.  For
this two-stage model, an analytical expression linking the burst
distribution to the steady-state distribution (analogous to
\eq{steady}) seems intractable. However, we can
derive expressions for steady-state mean and variance (see Appendix).  We
obtain:
\begin{equation}
	\bar{n}_s = \frac{\bar k}{\mu_p} \bar{n}_b,\label{mean-2}
\end{equation}

\begin{eqnarray}
	\frac{\sigma^2_s}{\bar{n}_s} &=&  1 + \bar{n}_b + \frac{1}{2}
    \left( \frac{\sigma^2_b}{\bar{n}_b} - (1 + \bar{n}_b) \right)
    \nonumber\\
	& &+ \left(1 + \frac{\lambda_{12}+\lambda_{21}}{\mu_p}\right)^{-1}
    \left(\frac{\lambda_{12}}{\lambda_{21}} \right)\left[ \frac{1-f}{1 + \frac{\lambda_{12}}{\lambda_{21}}f} \right]^2 \bar{n}_s,\label{var-2}
\end{eqnarray}

where we defined

\begin{equation}
	\bar k = \frac{\lambda_{21} k_1 + \lambda_{12} k_2}
    {(\lambda_{12}+\lambda_{21})}.
\end{equation}

Note that for the case $f=0$, the above formula reduces to previously
obtained results \cite{paulsson05,Shahrezaei2008}, whereas for $f=1$ we recover
\eq{var}. The above result thus generalizes  previously
obtained results for the case of nonzero $f$ and for arbitrary protein
burst distributions.

\section{Burst Distribution for Regulation by Small RNAs}\label{sect-sRNA}

The preceding sections derived general results connecting burst and
steady-state distributions for burst distributions which can deviate
from a geometric distribution. We now consider a specific regulation
scheme that can give rise to non-geometric protein burst
distributions: regulation by small RNAs. Small RNAs are genes that are 
transcribed but not translated, i.e. they are non-coding RNAs. In bacteria, 
small RNAs have been studied extensively in recent years \cite{Waters2009} 
in part due to the critical roles they play in cellular post-transcriptional 
regulation in response to environmental changes.

\begin{figure}[h!]
\begin{center}
\resizebox{100mm}{!}{
\includegraphics{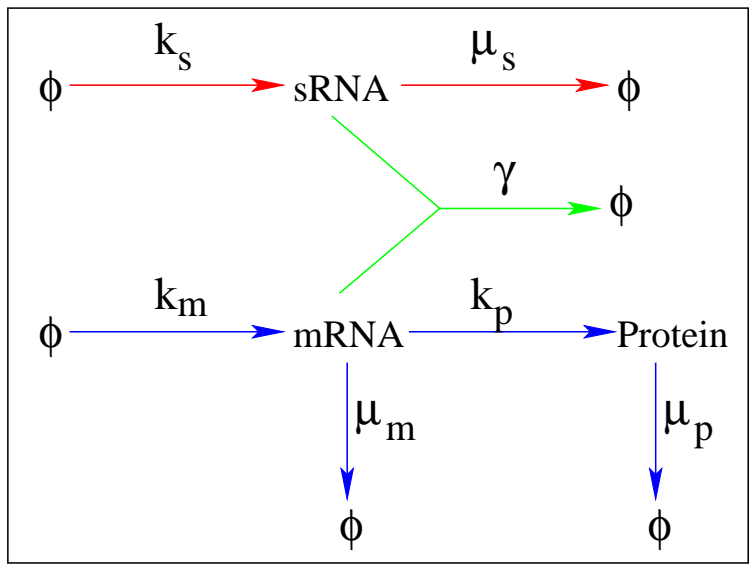}}
\caption{The sRNA-mRNA regulation scheme. The sRNA production rate
is $k_s$ and the degradation is $\mu_s$. The interaction rate between
the sRNA and mRNA that results in mutual degradation is $\gamma$. The
mRNA and Protein reaction rates are the same as shown in figure
\ref{figure_min_scheme}.}
\label{figure_sRNA_mRNA_target}
\end{center}
\end{figure}

The reaction scheme for
small RNA based regulation has been studied by several groups
\cite{levine07,mehta08,mitarai07,elgart_bj} and is schematically
represented in Figure \ref{figure_sRNA_mRNA_target}.
In the limit of large concentrations of the small RNA regulator, the fluctuations 
of the small RNAs can be neglected and a more general model can be analyzed \cite{jia10}.
However, when the fluctuations of the regulator cannot be neglected, the exact solution of 
the model represented in Figure \ref{figure_sRNA_mRNA_target} is analytically intractable 
and approximations schemes need to be employed. In the following, we show how, in the 
limit of infrequent protein bursts, an analytical expression for the generating function 
of the protein burst distribution can be derived which agrees well with simulations.

We consider the case that mRNA production is governed by a Poisson
process with constant rate $k_m$. In the limit of low $k_m$, the small
RNA distribution {\em prior} to each burst can be well approximated by
the unregulated small RNA distribution, which corresponds to a Poisson
distribution with mean $n_s = \frac{k_s}{\mu_s}$. With these
approximations, it is possible to derive an expression for the
regulated protein burst distribution due to interaction with small
RNAs as shown below.

Let us begin with the initial condition ($t=0$) corresponding to the
arrival of a mRNA. The protein burst distribution corresponds to the
number of proteins produced from this single mRNA until the time it is
degraded, either naturally or due to interaction with small RNAs. Our
approach will focus on first deriving an expression for the survival
probability of the mRNA at time $t$ ($S(t)$). Let us define $P_{1}(n,t)$ as 
the probability that the mRNA exists at time $t$ (i.e.\ it has not been 
degraded) and the number of sRNAs is $n$. Then, the mRNA survival probability
is given by $S(t) = \sum_{n=0}^\infty P_1(n,t)$.
Let us now define the operator $\hat H_s$ which acts as follows
\begin{equation}
       \hat H_s P(n) \equiv k_s\left[P(n-1) - P(n)\right]
       + \mu_s\left[(n+1)P(n+1) - nP(n)\right].\label{H_s}
\end{equation}
In terms of this operator, we can write down the Master equation for $P_{1}(n,t)$
as follows
\begin{equation}
	\partial_t P_1(n) = \hat H_s P_1(n) - \mu_m P_1(n) - \gamma n P_1(n),\label{p1}\\
\end{equation}
The corresponding initial condition is taken as
\begin{eqnarray}
	P_1(n,t=0) = e^{-n_{s}}\frac{n_{s}^{n}}{n!},\label{s-poisson}\\
\end{eqnarray}
where $n_{s}= (k_{s}/\mu_{s})$ (i.e., Poisson distribution of sRNAs at
time $t=0$) as discussed above.

In order to solve the \eq{p1} let us once again define a generating function
\begin{equation}
	G_1(x,t) \equiv \sum_{n=0}^\infty x^n P_1(n,t),
\end{equation}
which satisfies the partial differential equation
\begin{eqnarray}
	\partial_t G_1(x,t) = (\hat H_s - \mu_m - \gamma x\partial_x)G_1(x,t),\label{G_1}\\
	G_1(x,0) = \exp{(n_s(x-1))}.\label{s-ini}
\end{eqnarray}
Here the differential operator $\hat H_s$ can be easily derived from
the equation \eq{H_s}, namely $\hat H_s = (x-1)(k_s -
\mu_s\partial_x)$. The value of the generating function $G_1(x,t)$ at
point $x=1$ corresponds to $\sum_{n=0}^\infty P_1(n,t)$, i.e., the survival
probability $S(t)$ of the mRNA molecule at time $t$. This survival probability
can be obtained by solving \eq{G_1} using the method of
characteristics (Appendix). We obtain
\begin{equation}
	S(\tau) = \exp\left[-\alpha\left(1-e^{-\tau}\right) - \beta\tau\right],\label{survival}
\end{equation}
where we have defined the following dimensionless parameters:
\begin{equation}
	\tau = (\mu _s+\gamma)t, \label{const_tau}
\end{equation}
\begin{equation}
	\alpha = \left(n_{s} - \frac{k_{s}}{\mu_{s}+\gamma}\right)
        \frac{\gamma}{\mu _s+\gamma}, \label{const_alpha}
\end{equation}
\begin{equation}
	\beta = \frac{\mu_m}{\mu_s+\gamma} + \frac{\gamma k_s}
        {\left(\mu _s+\gamma \right)^2}, \label{const_beta}
\end{equation}

We can now proceed and calculate the generating function $G_b(x)$ of the
protein burst distribution. Since protein production occurs at a constant rate
$k_p$ during the mRNA lifetime, the number of proteins produced by a
surviving mRNA
in time $t$  is given by the Poisson distribution, with the corresponding
generating function given by $e^{k_p(x-1)t}$.
Since the difference $S(t)-S(t+\delta t)$ of
survival probabilities is the probability that the mRNA degrades within 
the time
interval $\{t,t+\delta t\}$, we obtain the burst generating function as
\begin{equation}
	G_b(x) = -\int_0^\infty\mathrm{d}t\, \partial_t S(t)e^{k_p(x-1)t}.
\end{equation}
Rewriting the burst size distribution in terms of dimensionless parameters 
results in the following integral form
\begin{eqnarray}
	G_b(x) = 1 - k (1 - x)\int_{0}^{1}\mathrm{d}z \,z^{k (1 - x) + \beta - 1}
	e^{\alpha (z - 1)},\label{burst-int}
\end{eqnarray}
where $k$ is yet another dimensionless parameter
\begin{equation}
	k \equiv \frac{k_p}{\mu_s+\gamma }.\label{const_k}
\end{equation}

\begin{figure}[h!]
\begin{center}
\resizebox{120mm}{!}{
\includegraphics{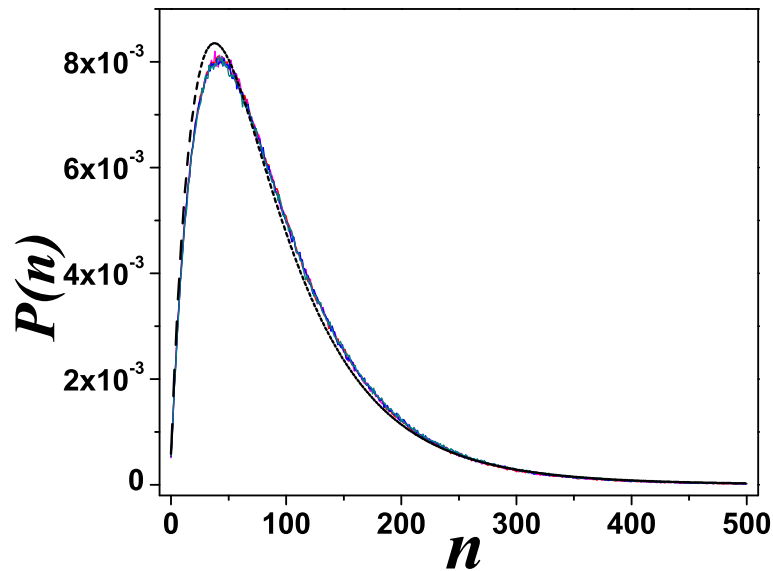}}
\caption{(color online) steady-state distributions with sRNA regulation. The
dashed curve corresponds to \eq{steady} using \eq{burst-int} with 
approximations (see text)
and $\alpha \simeq 4.76$, $\beta \simeq 1.34$, and $k \simeq 243.9$. The
other curves are the results from four sets of numeric simulations. See Table 1
for the values of the parameters used in the
simulations.}
\label{figure_sRNA_prob_small}
\end{center}
\end{figure}

The burst distribution with sRNA regulation, \eq{burst-int}, has some
interesting
features. We note that \eq{burst-int} predicts that the burst distribution
depends on three dimensionless parameters, $\alpha$, $\beta$,
and $k$ (equations (\ref{const_alpha}), (\ref{const_beta}), and
(\ref{const_k})) and the steady-state distribution (see \eq{steady}) only
adds a dependence on $k_m/\mu_p$. Thus the modulation of any of the
kinetic parameters shown in Figure \ref{figure_sRNA_mRNA_target} (for fixed 
$k_m/\mu_p$) should result
in the same steady-state distribution so long as the modifications
occur in such a way that $\alpha$, $\beta$, and $k$ remain
constant (and model assumptions/approximations are valid). 
As shown in Table 1, we can choose very different kinetic
parameters that give rise to the same values for $\alpha$, $\beta$
and $k$ and the prediction is that the burst and steady-state
distributions for these different parameter choices should collapse
onto a single curve.
To test this scaling prediction, we carried out
stochastic simulations based on the Gillespie algorithm
\cite{gillespie77} for a range of parameters such that $\alpha$,
$\beta$, and $k$ remain constant. From the simulations, we
recorded the resulting steady-state distributions and compared it to
the analytic result (see figure \ref{figure_sRNA_prob_small}).
For the choice of parameters noted, we observed that the burst distribution
is close to and can be well fitted by a geometric distribution. For a geometric
distribution, the steady-state and burst generating functions are related by
$G_{s}(x) = (G_{b}(x))^{\frac{k_b}{\mu_m}}$. We used this approximation to 
obtain
the analytical form of the steady-state generating function and derived the
steady-state protein distribution $P_{s}(n)$ using this. As can be seen in
Fig. \ref{figure_sRNA_mRNA_target}, the corresponding analytical results
are in good agreement with results from simulations. The simulation results 
are also consistent with the scaling prediction since the curves with 
different  parameter choices all collapse onto a single curve. The small 
discrepancy between the theoretical results and simulations is attributed to 
the approximations made, specifically the approximation for $G_{s}(x)$ noted above
which is strictly valid only if the burst distribution is geometric.  
The results 
obtained from simulations of  individual bursts are in 
very good agreement with the corresponding theoretical predictions.

\begin{table}
\begin{tabular}{|c|c|c|c|c|}
  \hline
  Simulation \# & $k_p$ & $k_s$ & $\mu_s$ & $\gamma$  \\
  \hline
  1 & 250 & 0.400313 & 0.072619 & 0.952381  \\
  2 & 300 & 0.717708 & 0.122308 & 1.107690  \\
  3 & 400 & 1.378120 & 0.217778 & 1.422222  \\
  4 & 500 & 2.055630 & 0.310870 & 1.739130  \\
  \hline
\end{tabular}
  \label{table_params}
  \caption{The values of the parameters used in the numeric simulations
  shown in figure \ref{figure_sRNA_prob_small}. For all simulations, $\alpha \simeq 4.76$, $\beta \simeq 1.34$, and $k \simeq 243.9$. Also,
  $\mu_m=1$, $k_m=0.01$, $\mu_p = 0.005$.}
\end{table}

\section{Summary}

Recent experiments underscore the need for connecting observed protein
distributions from single-cell and single molecule studies using
coarse-grained models of stochastic gene expression. In this context,
several results have been derived in the present study which will help
in the analysis of experimental results. We have shown how the
functional form of the underlying mRNA burst distributions can be
determined from observed protein distributions. If the protein burst
distribution is geometric then the corresponding mRNA burst
distribution has to be a conditional geometric distribution.  The
derived results further show that if the promoter can be repressed
such that observed protein bursts arise from single mRNAs, then the
underlying mRNA burst distribution in the unrepressed state can be
completely determined. Furthermore, we derive relations connecting means and variances for
burst and steady-state distributions for burst distributions which can
deviate from a geometric distribution.  The results derived also
provide insight into regulation of protein expression bursts by small
RNAs. The general results derived in this work can thus be used for
analysis of a wide range of models of gene expression. \\

The authors acknowledge funding support from  NSF (PHY-0957430) and from ICTAS,
Virginia Tech.


%
%

%
%

\section{Appendix}

\subsection{Relationship between mRNA and protein burst generating functions}

Let us define $P(m,n;t)$ as the probability to find $m$ mRNAs and $n$ 
proteins after time $t$ elapses since burst arrival.
The corresponding generating function $G_p(x,y;t) \equiv \sum_{m,n}x^m y^n P(m,n;t)$ satisfies the following partial differential equation:
\begin{equation}
	\der{t}G = \mu_m (1-x)\der{x}G + k_p(y-1)x\der{x}G.
\end{equation}
The equation above can be easily solved by the method of characteristics
\begin{equation}
	G(x,y;t) = G_m\left[\frac{1 - (1 + Y x)e^{-\mu_m Y t}}{Y}\right],
\end{equation}
where $G_m[.]$ is generating function of mRNAs at $t=0$, and we defined
\begin{equation}
	Y \equiv 1-\frac{k_p}{\mu_m}(y-1).
\end{equation}

Therefore, the time dependent distribution of proteins in the burst is given by generating function
\begin{equation}
	G_b(y;t) = G(1,y;t) = G_m\left[\frac{1 - (1 + Y)e^{-\mu_m Y t}}{Y}\right],
\end{equation}
and the corresponding steady state is simply
\begin{equation}
	G_b(y) = G_m\left[\frac{1}{Y}\right],
\end{equation}
which is identical to the equation in the main text (\eq{gen-mp}).

\subsection{Two stage model}

Assume that the mRNA production rate $k_m$ 
has a value $k_1$ in the state $1$ and a value $k_2$ in the state $2$. The 
state $1$ switches
with probability $\lambda_{12}$ into the state $2$ and back with probability $\lambda_{21}$. One gets
the following set of the equations for the generating functions 
$G_{1(2)}(x,t)$:
\begin{eqnarray}
	\partial_t G_1 = k_1 \left[G_b(x)-1\right]G_1 + \mu_p(1-x)\partial_x G_1 - \lambda_{12} G_1 + \lambda_{21} G_2,\\
	\partial_t G_2 = k_2 \left[G_b(x)-1\right]G_2 + \mu_p(1-x)\partial_x G_2 + \lambda_{12} G_1 - \lambda_{21} G_2.
\end{eqnarray}
Let us explicitly calculate two moments of the protein's  steady-state 
distribution. By setting
$x=1, t\rightarrow\infty$ we get
\begin{eqnarray}
	\lambda_{12} P_1^s = \lambda_{21} P_2^s,\\
	P_1^s + P_2^s = 1,
\end{eqnarray}
where $P_1^s\equiv G_{1}(1,\infty)$ and $P_2^s\equiv G_{2}(1,\infty)$ are 
steady-state probabilities 
to be in the states $1$ and $2$ accordingly. Therefore, one derives
\begin{eqnarray}
	P_1^s = \frac{\lambda_{21}}{\lambda_{12}+\lambda_{21}},\\
	P_2^s = \frac{\lambda_{12}}{\lambda_{12}+\lambda_{21}}.
\end{eqnarray}

By evaluating the first derivative with respect to $x$ at point $x=1$ one can 
calculate 
$\ave{n_{i}} \equiv \sum_{n=0}^{\infty} n P_{i}(n),\,i=1,2$:
\begin{eqnarray}
	0 = k_1 \ave{n_b}P_1^s - \mu_p \ave{n_1} - \lambda_{12} \ave{n_1} + \lambda_{21} \ave{n_2},\\
	0 = k_2 \ave{n_b}P_2^s - \mu_p \ave{n_2} + \lambda_{12} \ave{n_1} - \lambda_{21} \ave{n_2}.
\end{eqnarray}

Similarly, by evaluating the second order derivative with respect to $x$ at 
point $x=1$ one obtains
\begin{eqnarray}
	0 = k_1 \left[v_{b}P_1^s + 2\ave{n_b} \ave{n_1}\right] - 2\mu_p \lambda_1 - \lambda_{12} v_1 + \lambda_{21} v_2,\label{v1}\\
	0 = k_2 \left[v_{b}P_2^s + 2\ave{n_b} \ave{n_2}\right] - 2\mu_p \lambda_2 + \lambda_{12} v_1 - \lambda_{21} v_2,\label{v2}
\end{eqnarray}
where we defined 
\begin{eqnarray}
	v_{i} \equiv \sum_{n=0}^{\infty} n(n-1)P_{i}(n),\quad i=1,2\\
	v_{b} \equiv \sum_{n=0}^{\infty} n(n-1)P_{b}(n).
\end{eqnarray}

Hence, the average number of proteins in the steady-state and the variance 
can be derived by solving equations \eqs{v1}{v2} (result is given by 
the expressions \eqs{mean-2}{var-2} in the main text.)

\subsection{Derivation of survival probability for small RNA based regulation}

Solution of the equation \eq{G_1} using method of characteristics is given by
\begin{equation}
	G_1(x,t) = \exp\left[-\beta \tau + \frac{k_s(x-1)}{\gamma +\mu_s} +
	\frac{k_s\gamma }{\left(\gamma +\mu _s\right)^2}\right] g(z),
\end{equation}
where $\beta$ and $\tau$ are dimensionless parameters as defined in the main 
text and the function $g(z)$ needs to be determined from the initial 
condition \eq{s-ini}. Its 
argument is given by
\begin{eqnarray}
	z = \left[(x-1) + \frac{\gamma}{\gamma + \mu_s}\right]e^{-\tau}.
\end{eqnarray}

By matching the initial condition one gets
\begin{equation}
	g(z) = \exp\left[-\frac{k_s}{\gamma +\mu_s}z\right]
	\exp\left[n_s z -\frac{\gamma n_s}{\gamma + \mu_s}\right].
\end{equation}
Finally, since we are interested in the quantity $S(t) \equiv G_1(1,t)$ (survival probability), 
we obtain 
\begin{eqnarray}
        z\rightarrow \frac{\gamma }{\gamma + \mu_s}e^{-\tau},\\
	S(t) = \exp\left[-\beta \tau + \frac{k_s\gamma }{\left(\gamma +\mu _s\right)^2}\right] g(z).
\end{eqnarray}
from which the equation \eq{survival} from the main text can be obtained.\\

%
%
%
%


\bibliographystyle{unsrt}
\bibliography{stochastic_modeling}

\end{document}